\documentclass[a4paper, 11pt, reqno]{amsart}

\usepackage{mleftright}
\usepackage[utf8]{inputenc}
\usepackage[english]{babel}
\usepackage[T1]{fontenc}
\usepackage[left=3cm,right=3cm,top=3cm,bottom=3cm]{geometry}
\usepackage{amsmath}
\usepackage{float}
\usepackage{booktabs}
\usepackage{amssymb}
\usepackage{amsfonts}
\usepackage{mathtools}
\usepackage{latexsym}
\usepackage{graphicx}
\usepackage{pdflscape}
\usepackage{textcomp}
\usepackage{gensymb}
\usepackage{trsym}
\usepackage{url}
\usepackage{amsthm}
\usepackage{tikz-cd}
\usepackage{hyperref}
\usepackage{environ}
\usepackage[many]{tcolorbox}

\usepackage{cleveref}

\theoremstyle{remark}

\newtcbtheorem[use counter=prop, number within=section, crefname={definition}{definition}, Crefname={Definition}{definition}]
{definition}{Definition}{
	breakable,
	fonttitle = \bfseries,
	colframe = gray!75!black,
	colback = gray!10
}
{def}

\newtcbtheorem[use counter=prop, number within=section, crefname={theorem}{theorem}, Crefname={Theorem}{theorem}]
{theorem}{Theorem}{
	breakable,
	fonttitle = \bfseries,
	colframe = gray!75!black,
	colback = gray!10
}
{thm}

\newtcbtheorem[use counter=prop, number within=section, crefname={lemma}{lemma}, Crefname={Lemma}{lemma}]
{lemma}{Lemma}{
    breakable,
    enhanced,
    fonttitle = \bfseries,
    colframe = gray!5!black,
    colback = gray!15
}
{lem}

\tcbset{boxrule=1.4pt,boxsep=1pt,top=2pt,bottom=1pt}

\definecolor{mygreen}{RGB}{0, 104, 0}

\DeclarePairedDelimiterX\braket[2]{\langle}{\rangle}{#1 , #2}

\newcommand{\nnorm}[1]{
	{\left\vert\kern-0.25ex\left\vert\kern-0.25ex\left\vert #1
		\right\vert\kern-0.25ex\right\vert\kern-0.25ex\right\vert}
}

\newcommand{\defgr}{\mathrel{\mathop:\!\!=}}

\newcommand{\R}{\mathbb{R}}

\newcommand{\mc}{\mathcal}

\begin{document}

\title[Semiclassical formulae for Wigner distributions]{Semiclassical formulae for Wigner distributions}
\author{Sonja Barkhofen and Philipp Schütte and Tobias Weich}

\address{Institute for Photonic Quantum Systems, Paderborn University, Paderborn, Germany}
\email{sonja.barkhofen@uni-paderborn.de}
\email{pschuet2@mail.uni-paderborn.de}
\email{weich@math.uni-paderborn.de}

\begin{abstract}
In this paper we give an overview over some aspects of the modern mathematical theory of Ruelle resonances for chaotic, i.e. uniformly hyperbolic, dynamical systems and their implications in physics. First we recall recent developments in the mathematical theory of resonances, in particular how invariant Ruelle distributions arise as residues of weighted zeta functions. Then we derive a correspondence between weighted and semiclassical zeta functions in the setting of negatively curved surfaces. Combining this with results of Hilgert, Guillarmou and Weich yields a high frequency interpretation of invariant Ruelle distributions as quantum mechanical matrix coefficients in constant negative curvature. We finish by presenting numerical calculations of phase space distributions in the more physical setting of 3-disk scattering systems.
\end{abstract}

\maketitle

	
	

\begin{quote}
``Resonances of the time evolution (Perron-Frobenius) operator $\mathcal P$ for phase space densities have recently shown to play a key role for the interrelations of classical, semiclassical and quantum dynamics.''

\flushright{-- Weber-Haake-Seba (PRL85(17), 2000)}
\end{quote}
\vspace{5mm}

To any mathematician of that time this profound statement, made by Weber, Haake and Seba, must have sounded out of reach of rigorous mathematical analysis. Around the turn of the century there did not even exist a mathematically well-founded definition of the mentioned resonances for any physically realizable systems. However, today -- two decades later -- part of this vision has been transformed into rigorous mathematical theorems. The aim of the present article is to explain how these mathematical advances shed new light on semiclassical residue formulae for quantum states.

\section{Introduction} \label{intro}

One of the main paradigms of quantum chaos states that there exists a close and intricate relation between the periodic orbits of a chaotic classical dynamical system and the spectral data of its corresponding quantum system.
This topic has, since its very beginning, benefited vitally from an exchange between mathematics and physics. Visionary observations about physical systems like the quote above have stimulated rigorous mathematical theorems, while progress in terms of the precise mathematical understanding of the theoretical underpinnings enabled more and more profound observations in physics.\footnote{The basis for such a mutual stimulation is of course a common language and a good knowledge of the advances in both fields. With this article we plan to contribute to precisely this endeavor.} 
One example of such a bidirectional stimulation in quantum chaos starts with a remarkable mathematical result from representation theory:
Selberg's trace formula \cite{Sel56} can be interpreted in such a way that it gives an exact relation between eigenvalues of the Laplace-Beltrami operator and the periodic orbits of the geodesic flow on compact surfaces of constant negative curvature\footnote{
Such surfaces of negative curvature can be considered as mathematical models of quantum chaos: The geodesic flow (i.e. the classical free motion of a single particle on the surface) is always uniformly hyperbolic (i.e. chaotic) making the Laplacian (i.e. the Hamiltonian of a free quantum particle) a model for quantum chaos.}. 
Subsequently, Gutzwiller \cite{Gut71} has managed to show that such trace formulae can be derived as a semiclassical approximation in a much larger class of quantum mechanical systems. This in turn motivated Colin de Verdi\`ere \cite{CdVer73}, Chazarain \cite{Cha74}, and Duistermaat-Guillemin \cite{DG75} to prove the corresponding mathematical theorems which have been a central tool in spectral geometry ever since. Another example for a fruitful exchange between mathematics and physics in quantum chaos are the fractal Weyl bounds for open systems that have their roots in mathematical works of Sjöstrand \cite{Sjo90} and were transferred successfully into theoretical \cite{LSZ03, ST04} and experimental \cite{PWB+12} physics one and two decades later, respectively.

The relation between quantum spectrum and periodic orbits which can be deduced directly from the semiclassical trace formula states that the correlations in the quantum spectrum determine (after passing to the semiclassical limit) the length spectrum of the closed classical trajectories. The inverse question, i.e. to what extent one can determine the quantum spectrum purely from data of the classical periodic orbits, is much more subtle: For the very special case of constant curvature surfaces the exact Selberg trace formula allows for the construction of a Selberg zeta function whose zeros are in exact correspondence with the Laplacian's, i.e. quantum, spectrum (see \cite{Sel56,Hub59,Buser92,Bor16} for compact surfaces and \cite{PP01, BJP05, BO99} for the much more technically involved case of open dynamics).
For more general chaotic systems such a task is severely more complicated due to the semiclassical remainder terms in Gutzwiller's trace formula: Taking into account only the leading contributions one can formally define the so-called Gutzwiller-Voros zeta function \cite{Vor88}. Under the implicit assumption that large time and high frequency limits can be interchanged one can derive a relation between the zeros of the Gutzwiller-Voros zeta function and the quantum spectrum \cite[Eq.~(18) and related discussions]{Vor88}.
This relation has also been verified numerically with an impressive accuracy for certain special systems \cite{CE89,TSB+91,AF94, Wir99}.
Based on the semiclassical formulae for quantum eigenvalues it has also been possible to derive semiclassical formulae for the quantum eigenstates \cite{Sie07} as well as for their phase space distributions \cite{AF93, EFMW92}.
A mathematically rigorous understanding of the correspondence between zeros of the Gutzwiller-Voros zeta function and the quantum spectrum, however, is still out of reach (c.f. discussion in \cite[Section~3.3.1]{FT17a}).

A related field in which significant mathematical progress has been achieved over the past two decades is the spectral theory of Ruelle resonances.
Nowadays, this theory is reasonably well understood and allows the definition of a discrete resonance spectrum associated with a chaotic or, more precisely, uniformly hyperbolic classical dynamical system.
This theory comes with exact trace formulae which again enable the construction of zeta functions that have exact spectral interpretations. 

With these remarks in mind, the present article aims to explain how only recently introduced weighted zeta functions for Ruelle resonances are related to semiclassical residue formulae for quantum phase space distributions.
In general, these weighted zeta functions allow us to calculate classical Ruelle resonant states. 
In special cases (again negatively curved surfaces or more generally rank one locally symmetric spaces) it can, however, even be proven that the residues yield phase space distributions of quantum states.
In order to embed the recent mathematical findings into the context of physical systems we will explain how these rigorous results are related to semiclassical residue formulae which had already been derived by Eckhardt et al. in the early 90s \cite{EFMW92}.
These relations shed new light on these formulae. Furthermore we demonstrate that the residues can be calculated numerically for 3-disk systems in an efficient manner. 
We compare the numerical results obtained by the classical residue formulae with phase space distributions of a full quantum calculation and obtain a good agreement on a qualitative level. 

\subsection{Paper organization}

In Section~\ref{sec:Ruelle} we give a brief introduction to the theory of Ruelle resonances, we define and discuss shortly the invariant Ruelle distribution associated with any Ruelle resonance, and describe a recent result on how invariant Ruelle distributions arise as residues of certain weighted zeta functions. This lays a foundation for the entire remainder of this article. In Section~\ref{sec:semiclassical} we recall the definition of weighted semiclassical zeta functions and show how this can be related to weighted zeta functions for invariant Ruelle distributions in the model case of compact surfaces of negative curvature. Specifying the model case even further to spaces of constant negative curvature we discuss in Section~\ref{sec:exact_qcc} a particular instance of the quantum classical correspondence for locally symmetric spaces and how it relates classical with quantum mechanical phase space distributions. In Section~\ref{sec:numerics} we finally combine most of the material introduced in the previous chapters to be able to present and interpret numerical results on classical as well as quantum mechanical phase space distributions for 3-disk scattering systems, all of which are calculated with the aid of certain zeta functions. In particular, we compare quantum mechanical with classical resonances/residues and demonstrate how semiclassical techniques open a high frequency domain inaccessible to fully quantum mechanical calculations.

\section{Ruelle resonances, resonant states and weighted zeta functions} \label{sec:Ruelle}

Let $\mathcal P$ be the phase space of a classical physical system modeled by a symplectic manifold, and $H:\mathcal P\to \R$ a classical Hamiltonian with a regular energy value $E\in\mathbb{R}$. Then the energy shell $\mathcal M := H^{-1}(E)\subset \mathcal P$ is a smooth submanifold. The classical dynamics is conveniently described by Hamilton's equations on $\mathcal{M}$, i.e. is given by a flow which we denote $\varphi_t$ and which in turn is generated by a vector field $X$. Note that the Liouville measure provides a canonical flow invariant measure on the energy shell $\mathcal M$ and the time evolution of a classical probability density $\rho$ gets controlled by the transfer operator $\exp(-Xt)$, explicitly given by 
\[
 e^{-Xt}\rho = \rho\circ\varphi_{-t}.
\]

If the differential operator $-X$ had discrete spectrum $\sigma(-X)$ such that $\lambda = 0$ was a simple eigenvalue and every other eigenvalue was contained in some left halfplane, i.e. $\sigma(-X)\setminus \{0\} \subseteq \{\lambda\in \mathbb{C} \,|\, \mathrm{Re}(\lambda) < -\gamma < 0\}$, then the time evolution of $\rho$ would converge exponentially to an equilibrium distribution described by the eigenstate of the leading eigenvalue at zero. The set of (generalized) eigenvalues of $X$ that describe such a convergence to equilibrium are called Ruelle (or sometimes Pollicott-Ruelle) resonances due to pioneering works of Ruelle \cite{Rue76,Rue87b, Rue89} and Pollicott \cite{Pol85} in the 1980s. The general idea of studying discrete spectra for Liouville operators such as $\exp(-Xt)$ has also been used successfully in the physics literature \cite{GR89,GR92} to tackle questions regarding convergence to equilibrium for chaotic dynamical systems. In particular, we mention here the work by Haake and coauthors \cite{WHS00,WHB+01, MWH01, OMZH03}.  

The main difficulty in rigorously defining these Ruelle resonances lies in the fact that the ordinary $L^2$-spectrum of $X$ usually includes continuous portions equal e.g. to the entire imaginary axis. In order to make sense of a discrete resonance spectrum one needs to construct particular Hilbert or Banach spaces as domains for $X$ on which the spectrum becomes discrete. The prototype of such a theory has been delivered by Ruelle \cite{Rue89} for expanding maps, but it took another 13 years until it was understood how to construct these spaces for general hyperbolic diffeomorphisms \cite{BKL02} and even longer to treat hyperbolic flows. Today one has the following result:

\begin{theorem}{Existence of Ruelle resonances \cite{BL07, FS11, DZ16a}}{exist}
Let $X$ be the vector field of a uniformly hyperbolic flow on a compact manifold $\mathcal M$. Then the resolvent $\mathrm{R}(\lambda) \defgr (-X - \lambda)^{-1}$ is well defined and holomorphic for $\textup{Re}(\lambda) > 0$ as an operator on $\mathrm{L}^2(\mathcal M)$ and has a unique extension to a meromorphic family of continuous operators $\mathrm{R}(\lambda): \mathrm{C}^\infty(\mathcal M) \to \mathcal{D}'(\mathcal M)$.
\end{theorem}

The discrete set of poles of the extension $\mathrm{R}(\lambda)$ is then defined to be the set of \emph{Ruelle resonances}. 
In particular they arise as an actual discrete spectrum in special Hilbert spaces and those Hilbert spaces in fact provide the key ingredient in the proof of the meromorphic continuation.
As the construction of these spaces is technically somewhat demanding we refrain from entering into the details and refer to the works cited above.

We would like to mention, however, that for a given resonance $\lambda_0$ the residue operator of $\mathrm{R}(\lambda)$ at $\lambda_0$ has finite rank and yields a projector onto the space of generalized eigen-/resonant states. More precisely, if $\mathcal D_{E_u^*}'(\mathcal M)$ is the set of distributions with Hörmander wavefront set contained in $E_u^* \defgr (E_0\oplus E_u)^\perp$ (with $E_0\oplus E_s\oplus E_u$ being the splitting into neutral, stable and unstable bundle associated with the flow) then the residue at $\lambda_0$ is a projector as an operator $\mathcal D_{E_u^*}'(\mathcal M)\to \mathcal D_{E_u^*}'(\mathcal M)$ \cite[Section 2.3.2]{CDKP20} onto the space
\begin{equation*}
\{u\in \mathcal D'_{E_u^*}(\mathcal M), (-X-\lambda_0)^Ju = 0 \text{ for some } J>0\} .
\end{equation*}

For a physics oriented introduction to the wavefront set we refer to \cite{BDH14}. For the further understanding of this article it is, however, sufficient to simply think of $\mathcal D'_{E_u^*}(\mathcal M)$ as a space containing all distributions that are smooth in the weak unstable directions given by $E_0 \oplus E_u$.

Note that generically one can assume that both the order of a pole at a resonance and the rank of the associated residue operator are equal to one. 
Under this assumption and in a physicist's notation one would write the residue operator as $\Pi_{\lambda_0} = \vert u\rangle \langle v\vert$, where $(-X - \lambda_0) \vert u\rangle = 0$ and $\langle v\vert (-X - \lambda_0) = 0$ are right and left eigenvectors, respectively. 
Writing $M_a$ for the multiplication operator with a smooth function $a\in \mathrm{C}_\mathrm{c}^\infty(\mathcal M)$ we will be particularly interested in the following distribution which we call the \emph{invariant Ruelle distribution associated with $\lambda_0$}\footnote{To be precise this defines a distribution after fixing a smooth Lebesgue measure on $\mathcal M$. 
For a physical system one conveniently chooses the Liouville measure. 
However, the definition is also possible without fixing a measure -- in this case one obtains a generalized density. 
An even more general definition without assumptions on the pole order or the rank of the projector exists. We refrain from presenting it to avoid technical complications and refer to \cite{BSW21} for details.}:
\[
 \mathcal T_{\lambda_0}: \mathrm{C}^\infty_\mathrm{c}(\mathcal M)\ni a\longmapsto \mathcal{T}_{\lambda_0}(a) \defgr \langle v\vert M_a\vert u\rangle = \textup{Tr} \left( \Pi_{\lambda_0} M_a \right) \in \mathbb{C} .
\]

At first glance the product of distributions $u, v$ written above seems problematic. But it has been shown \cite{FS11,DZ16} that $v\in\mathcal D'_{E_s^*}(\mc M)$ and thus the above expression is well defined by transversality of the wavefront sets of the two distributions. A simple calculation now yields
\[
 \mathcal T_{\lambda_0}(Xa)= \langle v|M_{Xa}|u\rangle = \langle v\vert [X + \lambda_0, M_a] \vert u\rangle = 0 ~,
\]
which implies that the distribution $\mathcal T_{\lambda_0}$ is in fact invariant under the flow, i.e. $X\mathcal T_{\lambda_0} = 0$. In summary we thus associated with any Ruelle resonance an invariant distribution on the energy shell $\mc M$ in a very natural way.\footnote{In order to underline the canonical nature of $\mathcal{T}_{\lambda_0}$ let us draw an analogy to quantum mechanics:
The spectral projector of a simple quantum eigenvalue is given by $\vert\psi\rangle \langle\psi\vert$ and proceeding in analogous fashion as above one obtains the distribution $|\psi|^2$.} 
For the leading resonance this distribution coincides with a well-known and important invariant measure called SRB- or, in presence of additional potentials, Gibbs measure and has been studied extensively in the dynamical systems community \cite{You02}. 
It has recently been proven that the invariant Ruelle distributions appear as residues of weighted zeta functions in the following manner:

Given $a\in\mathrm{C}^\infty(\mathcal{M})$ and $\lambda\in \mathbb{C}$ we define
\begin{equation} \label{eq_zeta_def}
Z_a^\text{Ruelle}(\lambda) \defgr \sum_{\gamma} \left( \frac{\exp\left(-\lambda L_\gamma\right)}{\vert \det(\mathrm{id} - \mathcal{P}_\gamma) \vert} \int_{\gamma^\#} a \right) ,
\end{equation}
where the sum extents over all closed trajectories $\gamma$ of $\varphi_t$, $L_\gamma$ is the period of $\gamma$, $\gamma^\#$ denotes the corresponding primitive closed trajectory, and $\mathcal{P}_\gamma$ its linearized Poincar\'{e} map. 
It can then be proven \cite[Theorem~1.1]{BSW21} that the weighted zeta function converges absolutely for $\textup{Re}(\lambda)\gg 0$ to a holomorphic function in this region and that this function can be continued meromorphically to the whole plane. Any pole $\lambda_0$ of this function is then a Ruelle resonance and the residue is given by
\begin{equation}\label{eq:ruelle_residue}
\textup{Res}_{\lambda = \lambda_0} \left[Z^\text{Ruelle}_a(\lambda) \right] = \mathcal T_{\lambda_0}(a) = \langle v \vert M_a \vert u\rangle ,
\end{equation}
where the first equality holds in full generality and the second under the multiplicity one assumption made above. It was furthermore demonstrated in \cite[Appendix~A]{BSW21} that this relation between invariant Ruelle distributions and residues provides an approach for efficient numerical calculations by applying cycle expansion to the weighted zeta functions. 

Note that so far we discussed the case of a uniformly hyperbolic flow on a compact energy shell, i.e. a closed system. 
All the above statements also hold (sometimes with slight modifications) for open systems with uniformly hyperbolic dynamics on a compact trapped set.
We refer to \cite{DG14, BSW21} for the necessary technical assumptions and precise statements of the results.

Finally the statements can be generalized to (complex) vector bundles $\mathcal{E}$ over the energy shell. The corresponding dynamical data is provided by a first-order differential operator $\mathbf{X}$ acting on sections of $\mathcal{E}$. 
The theorems described above continue to hold with $X$ replaced by $\mathbf{X}$ and some additional minor modifications\footnote{Such as replacing the spaces of distributions with spaces of vector-valued distributions.} if the following Leibniz rule connecting dynamics in the bundle with dynamics on the base holds:
\begin{equation*}
\mathbf{X}(a\cdot \mathbf{u}) = a\cdot \mathbf{X}\mathbf{u} + (X a)\cdot \mathbf{u}, \qquad a\in \mathrm{C}^\infty(\mathcal{M}),\, \mathbf{u}\in \mathrm{C}^\infty(\mathcal{M}, \mathcal{E}) .
\end{equation*}
A particular instance of this setup will be of importance in the next section.

\section{Semiclassical residue formulas}\label{sec:semiclassical}

As mentioned in the introduction there are several theoretical works that derive periodic orbit formulae for quantum eigenstates or related quantities such as quantum phase space distributions. Let us recall the results by Eckhardt et al. \cite{EFMW92} (see also \cite[Section~5]{Sie07}): Let $\hat H$ be a Hamiltonian operator, $\vert n\rangle$ an eigenbasis for this Hamiltonian, and $\widehat{A}$ a quantum observable whose classical counterpart we denote by $a$. Built on an extension of Gutzwiller's work the authors then introduce a weighted zeta function:\footnote{Note that $Z^\text{sc}_a$ is, up to normalization, denoted by $R_{a,osc}$ in \cite{EFMW92}, see Eq.~(24) therein. Our formula actually differs from the definition in \cite{EFMW92} by a factor of $-\mathrm{i}$ but presumably this is a typo in the corresponding reference as otherwise the results in \cite{EFMW92} would contradict the fact that the Gutzwiller-Voros zeta function vanishes to the order of the multiplicity of the quantum eigenspace.}
\begin{equation}\label{eq:sc-zeta}
Z^\text{sc}_a(E):= -\mathrm{i} \sum_{\gamma \subset \Sigma_E} \frac{e^{\mathrm{i}S_\gamma(E)/h - \mathrm{i}\mu_\gamma\pi/2}}{|\det(1 - \mc P_\gamma)|^{1/2}} \left(\int_0^{T_{\gamma^\#}} a(\gamma(t))dt\right).
\end{equation}
Here $S_\gamma(E)$ is the classical action along the closed orbit $\gamma$ and $\mu_\gamma$ the Maslov index. As before $P_\gamma$ and $\gamma^\#$ denote the linearized Poincar\'e map and the primitive orbit corresponding to $\gamma$. Note that, however, in contrast to \eqref{eq:ruelle_residue} the periodic orbits are now those in the energy shell $\Sigma_E$ and thus they depend implicitly on the variable $E$.\footnote{Let us note a subtle point: If $E$ varies in $\mathbb{R}$ the $E$-dependence does not pose any problems. Often, however, one would like to consider meromorphic continuations of $Z^\text{sc}_a(E)$, e.g. when defining complex resonances for open systems. Then holomorphic continuation of $\int_{\gamma^\#} a$ might pose a severe problem and require strong assumptions on the Hamiltonian. From this point of view it might be advantageous to work with the zeta function $Z^\text{FT}_a$ defined later, where such problems do not occur.} The authors deduce that $Z^\text{sc}_a(E)$ has poles at the ``semiclassical eigenenergies'' and in a regime where such a semiclassical eigenenergy corresponds to a quantum eigenvalue $E_n$ they argue that in the semiclassical limit (c.f. \cite[Eq.~(35)]{EFMW92})
\begin{equation}\label{eq:sc_residue}
\langle n\vert \widehat{A}\vert n\rangle  \approx \textup{Res}_{E=E_n} \left[Z^\text{sc}_a(E) \right] .
\end{equation}

Thus the matrix coefficient of the quantum observable $\widehat{A}$ w.r.t. a quantum eigenstate $|n\rangle$ can be expressed as a residue of a weighted zeta function. The similarity between the semiclassical weighted zeta function \eqref{eq:sc-zeta} and the Ruelle weighted zeta function \eqref{eq_zeta_def} as well as between the residue formulae \eqref{eq:ruelle_residue} and \eqref{eq:sc_residue} is striking. We now intend to explain how the residues $\textup{Res}_{E=E_n} [Z^\text{sc}_a(E)]$ can indeed be brought into an \emph{exact} correspondence with the invariant Ruelle distributions by describing a setting in which the semiclassically approximate formula \eqref{eq:sc_residue} becomes a rigorous theorem.

We begin by assuming the mathematical model case of an oriented closed surface $N$ of negative sectional curvature. 
In this case the classical phase space is given by the cotangent bundle $T^*N$. We take the Hamiltonian $H(q,p) = |p|^2$ then the unit energy shell under consideration is the sphere bundle $SN$ and the corresponding quantum Hamiltonian $\hat H = \Delta$ the (positive) Laplace-Beltrami Operator.   
In this setting the Maslov index $\mu_\gamma$ of any closed geodesic $\gamma$ vanishes and the action is given by $S_\gamma(E) = hET_\gamma = h\sqrt{E} L_\gamma$ where $L_\gamma$ is the length of the closed orbit. We therefore conclude
\begin{equation*}
Z^\text{sc}_a(E) = -\mathrm{i}\sum_{\gamma\subset \Sigma_E} \frac{\exp(\mathrm{i}\sqrt{E}\cdot L_\gamma)}{|\det(1 - \mc P_\gamma)|^{1/2}} \left(\int_0^{T_{\gamma^\#}} a(\gamma(t))dt\right).
\end{equation*} 

Now inspired by the works of Faure and Tsujii \cite{FT15, FT17a, FT21} on the mathematical foundations of the Gutzwiller-Voros zeta function, we will lift the flow to the $1/2$-density bundle and show that the residues studied by Eckhardt et al. can be identified with invariant Ruelle densities of this lifted flow:
Therefore recall that in the present setting the stable bundle $E_s$ is a one-dimensional vector bundle and we consider the corresponding dual bundle of the $1/2$-density bundle denoted by $|E_s|^{-1/2}$. The geodesic flow on $SN$ lifts by differentials to $|E_s|^{-1/2}$. 
By a straight forward calculation based on \cite[Theorem~1.1]{BSW21}\footnote{\label{foot:hölder}We would like to emphasize a mathematically important subtlety. 
Theorem~1.1 in \cite{BSW21} only holds for smooth vector bundles whereas the bundle $|E_s|^{-1/2}$ possesses only Hölder regularity. 
Nonetheless it is plausible that \cite[Theorem~1.1]{BSW21} can be extended to the present setting by adapting the FBI techniques of  \cite{FT21} to the present setting. An even easier alternative could be to work with Grassmanian extension techniques from \cite{FT17a} instead of the density bundles. 
This has very recently even been transformed to the setting of obstacle scattering \cite{CP22}. 
Establishing a completely rigorous foundation for $Z^\text{FT}_a(\lambda)$ should be subjected to further mathematical research.} one obtains the following expression for the corresponding weighted zeta function\footnote{As the lift to the half density bundles $|E_s|^{-1/2}$ was intensively studied by Faure and Tsujii we call this version of the zeta function the Faure-Tsujii zeta function.}
\[
 Z^\text{FT}_a(\lambda) = \sum_{\gamma} \left( \frac{\exp\left(-\lambda L_\gamma\right) |\Lambda_\gamma|^{1/2}}{\vert \det(\mathrm{id} - \mathcal{P}_\gamma) \vert} \int_{\gamma^\#} a \right) ,
\]
and the residues of $Z^\text{FT}_a(\lambda)$ correspond to the invariant Ruelle distributions of the lifted geodesic flow. We can now compare the semiclassical and the Faure-Tsujii zeta functions: First of all the symplectic structure of $E_s\oplus E_u$ implies that for any closed geodesic $\gamma$ we can write
\begin{equation*} 
\mc P_\gamma =
\left(\begin{array}{cc}
\Lambda_\gamma & 0\\
0 & \Lambda_\gamma^{-1}
\end{array}\right) , \quad \text{ where } \Lambda_\gamma > 1 .
\end{equation*}
This lets us immediately deduce
\begin{equation}\label{eq:pdevel1}
\frac{\Lambda_\gamma^{1/2}}{|\det(1-P_\gamma)|} = \frac{\Lambda_\gamma^{1/2}}{\Lambda_\gamma\cdot (1-\Lambda_\gamma^{-1})^2} = \Lambda_\gamma^{-1/2} \sum_{n=0}^{\infty} (n + 1) \Lambda_\gamma^{-n} ,
\end{equation}
and furthermore
\begin{equation}\label{eq:pdevel2}
\frac{1}{|\det(1-P_\gamma)|^{1/2}}=\frac{1}{\Lambda_\gamma^{1/2}\cdot (1-\Lambda_\gamma^{-1})} = \Lambda_\gamma^{-1/2} \sum_{n=0}^{\infty} \Lambda_\gamma^{-n} .
\end{equation}
We can thus expand both weighted zeta functions in series over their independent variables $\lambda$ and $E$ by plugging these expansions into the respective definitions:
\[\begin{split}
 Z^\text{FT}_a(\lambda) = \sum_{n=0}^\infty Z^{\text{FT}, n}_a(\lambda) &~\text{ where }~ Z^{\text{FT}, n}_a(\lambda) =  \sum_{\gamma} \left( \exp\left(-\lambda L_\gamma\right) (n+1) \Lambda_\gamma^{-1/2-n} \int_{\gamma^\#} a \right) , \\
 Z^\text{sc}_a(E) = \sum_{n=0}^\infty Z^{\text{sc}, n}_a(E) &~\text{ where }~ Z^{\text{sc}, n}_a(E) =    -\mathrm{i}\sum_{\gamma\subset{\Sigma_E}} \left( \exp\left(\mathrm{i}\sqrt{E}\cdot L_\gamma\right) \Lambda_\gamma^{-1/2-n} \int_{\gamma^\#} a \right) .
 \end{split}
\]
Let us compare the individual terms in these expansions: Any periodic orbit $\gamma\subset \Sigma_E$ of the Hamiltonian $H(q,p)=|p|^2$ corresponds by rescaling to a unit speed geodesic $\tilde \gamma(t) = \frac{1}{\sqrt{E}}\cdot\gamma(t/(2\sqrt{E}))$ where the multiplication of $\gamma$ is understood in the fibers of $T^* N$.
Now, if we additionally assume that the observable $a$ is invariant under rescaling of the $p$-variable, i.e. $a(q, p)= a(q,\nu p)$ for $\nu, |p| > \varepsilon$, we obtain 
\[
\int_0^{T_{\gamma}}a(\gamma(t))dt= \frac{1}{2\sqrt{E}}\int_{\tilde \gamma} a ,
\]
and by comparing the leading terms in Equations \eqref{eq:pdevel1} and \eqref{eq:pdevel2} we directly observe the equality 
\begin{equation}\label{eq:compare_zetas}
Z^{\text{FT}, 0}_a(\lambda) = 2\mathrm{i}\sqrt{E} \cdot Z^{\text{sc}, 0}_a(E) \Big|_{E = (\mathrm{i}\lambda)^2} .
\end{equation}
Consequently (except for $\lambda=0$) the poles $\lambda_n$ of $Z^{\text{FT},0}_a(\lambda)$ are in one to one correspondence to the poles $E_n = (\mathrm{i}\lambda_n)^2$ of $Z^{\text{sc}, 0}_a(E)$ and one gets the following equality of residues
\begin{equation}\label{eq:compare_residues}
\textup{Res}_{\lambda=\lambda_n}(Z^{\text{FT}, 0}_a(\lambda))= \textup{Res}_{E=E_n}(Z^{\text{sc}, 0}_a(E)) .
\end{equation}

Furthermore the assumption of negative sectional curvature guarantees uniform hyperbolicity of the dynamics which in turn yields the existence of constants $C > 0$ and $\beta_{\mathrm{min}} > 0$ such that
\begin{equation}\label{eq:uniform_estimate}
\Lambda_\gamma^{-1} < C e^{-\beta_{\textup{min}}L_\gamma} .
\end{equation}
If we define the so-called \textit{topological entropy} to be 
\[
 h_\mathrm{top} = \lim_{L\to\infty} \frac{\log(\#\{\gamma: L_\gamma<L\})}{L} ,
\]
then one checks that $\sum_{n=1}^\infty Z^{\text{FT}, n}_a(\lambda)$ and $\sum_{n=1}^\infty Z^{\text{sc}, n}_a((\mathrm{i}\lambda)^2)$ converge uniformly for $\textup{Re}(\lambda)\geq h_\mathrm{top} - \frac{3}{2} \beta_\mathrm{min}$ and thus contribute to the respective complete weighted zetas with a holomorphic function in this right halfplane. In particular we have shown that for $\textup{Re}(\lambda)\geq h_\mathrm{top} - \frac{3}{2}\beta_\mathrm{min}$ the poles and residues are completely determined by $Z^{\text{FT},0}_a(\lambda)$ and $Z^{\text{sc},0}_a(E)$ and by \eqref{eq:compare_residues} the corresponding residues coincide. 

Let us summarize what we have shown by the above arguments and calculations in the special case of negative curvature surfaces: Any pole $\lambda_n$ of the semiclassical weighted zeta function $Z^\text{sc}_a\left((\mathrm{i}\lambda)^2\right)$ \emph{in the region $\textup{Re}(\lambda) > h_\mathrm{top} - \frac{3}{2} \beta_\mathrm{min}$} is a Ruelle resonance of the geodesic flow lifted to the bundle $|E_s|^{-1/2}$. Furthermore the corresponding residue is given by the associated invariant Ruelle distribution. With this in mind the semiclassical arguments presented in \cite{EFMW92} suggest that for any Ruelle resonance $\lambda_n$ of the geodesic flow lifted to $|E_s|^{-1/2}$ there exists a quantum eigenvalue near $(\mathrm{i}\lambda_n)^2 / 2$ and that the associated invariant Ruelle distribution is approximately related to a matrix coefficient of the quantum observable:
\begin{equation} \label{eq:compare_inv_ruelle_quant_matrix}
\mathcal{T}_{\lambda_n}(a) \approx \langle n|\hat A|n\rangle .
\end{equation}

\section{Exact quantum classical correspondence on surfaces of constant negative curvature} \label{sec:exact_qcc}

In this section we will summarize some recent mathematical results that establish an exact correspondence between Ruelle resonances and quantum eigenvalues on surfaces of \emph{constant} negative curvature. In particular we will discuss how this provides a rigorous analogue to \eqref{eq:sc_residue}.

For now we assume that the closed surface $N$ introduced in the previous section has \emph{constant} negative curvature. This is the simplest example of a Riemannian locally symmetric space of rank one and most of what is discussed here generalizes to this setting, with some aspects generalizing even further to the case of higher rank spaces \cite{HWW21}. Assuming this setup the following simplifications take place: 
\begin{equation}\label{eq:constant_curvature}
\Lambda_\gamma = e^{L_\gamma}, \quad h_\mathrm{top} = 1, \quad \beta_\mathrm{min} = 1 .
\end{equation}
At the level of resonances the constant curvature implies that lifting the geodesic flow to the line bundle $|E_s|^{-1/2}$ invokes nothing but a shift of the set of Ruelle resonances by $+1/2$. On the level of resonant states and invariant Ruelle distributions the lift acts trivially. During the last years it has been worked out in a series of mathematical papers that an exact relation between the Ruelle resonances and resonant states on constant curvature surfaces and the corresponding quantum eigenvalues and eigenstates holds. We summarize the results in a formulation suitable for the present article as follows:

\begin{theorem}{Quantum classical correspondence}{qcc}
Let $N$ be a closed surface of constant negative curvature. Then any Ruelle resonance $\lambda_n$ of the geodesic flow lifted to $|E_s|^{-1/2}$ with $\textup{Re}(\lambda_n) > -1/2$ corresponds to an eigenvalue $E_n = (\mathrm{i}\lambda_n)^2 + 1/4$ of the quantum Hamiltonian $\hat H = \Delta_N$. Given any sequence of Ruelle resonances $\lambda_j= \mathrm{i}r_j \in \mathrm{i}\R$ with $r_j \to \infty$ and any fixed $a\in \mathrm{C}^\infty(SN)$ one has
\[
 \mathcal T_{\lambda_j}(a) = \sum_{\phi_n} \langle\phi_n|\mathrm{Op}_{1/r_j}(a)|\phi_n\rangle + \mathcal{O}(1 / r_j) .
\]
Here the sum ranges over an orthonormal basis of the $\Delta_N$-eigenspace with eigenvalue $1/4 + r_j^2$ and $\mathrm{Op}_h(a)$ denotes some fixed semiclassical quantization of the classical observable $a$.
\end{theorem}

This theorem should be compared with  \eqref{eq:compare_inv_ruelle_quant_matrix} and the discussions in Section~\ref{sec:semiclassical}. Apart from the missing shift\footnote{Such shifts are ubiquitous in the analysis on locally symmetric spaces and are often called $\rho$-shifts.} of $1/4$, \Cref{thm:qcc} is thus a rigorous version of the semiclassical predictions made in \cite{EFMW92}. 

Let us give a short bibliographic summary of the development that led to \Cref{thm:qcc}: The relation between the classical and quantum spectra was shown in \cite{DFG15}, but see also \cite{FF03} for previous related work and \cite{GHW21, KW21} for generalizations. The relationship between invariant Ruelle distributions and matrix coefficients was proven in \cite{GHW21} based on previous work on so-called Patterson Sullivan distributions in \cite{AZ07, HHS12}. We would like to emphasize that all the above mentioned works heavily rely on Lie theoretic tools such as representation theory of Lie groups and Lie theoretic structure theory of Riemannian symmetric spaces. While those techniques provide powerful mathematical tools to prove results on locally symmetric spaces (such as constant curvature surfaces) we see no hope of transferring these methods to more general chaotic systems which possess fewer intrinsic symmetries.

\section{Numerical studies for 3-disk systems} \label{sec:numerics}

As explained in the previous section it is rigorously known that for locally symmetric spaces the invariant Ruelle distributions are related to quantum matrix coefficients.
For systems beyond this rather restrictive class of mathematical models of quantum chaos, the question about this relation is mathematically completely open. 
However, for many realistic physical systems both quantities are well defined.\footnote{Modulo the regularity problem regarding the bundle $|E_s|^{-1/2}$, see footnote \textsuperscript{\ref{foot:hölder}}.} In this section we will demonstrate that for 3-disk scattering systems we can numerically calculate both, the invariant Ruelle distributions as well as the quantum matrix coefficients, which allows us to compare them. 
As predicted by semiclassical analysis, we find a good qualitative agreement between the two objects. 

Let us give a short introduction to 3-disk systems (or more generally convex obstacle scattering). 
To this end consider the physical situation of a single particle moving freely in the 2-dimensional Euclidean plane and scattering at 3 non-intersecting disks whose centers are placed on an equilateral triangle.

\begin{figure}[H]
	\includegraphics[scale=0.7]{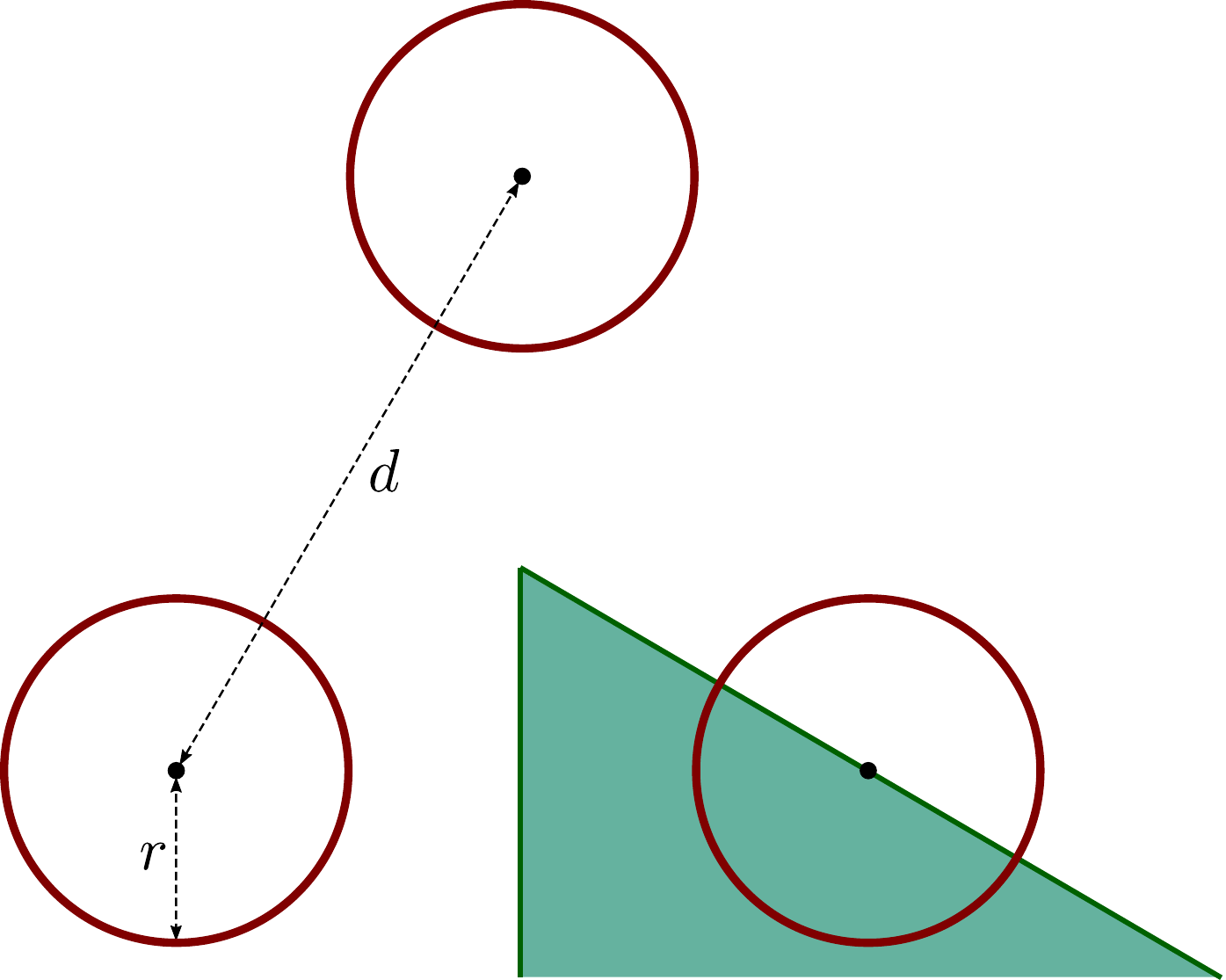}
	\caption{Illustration of a fully symmetric $3$-disk system given by its defining parameters $r$, the common disk radius, and $d$, the mutual distance between the disks. For our purposes such a 3-disk system is fully characterized by the ratio $d / r$. A fundamental domain for symmetry reduction, which is the model of choice for experimental realization, is given by the region shaded in green.}
	\label{fig:3disk}
\end{figure}

This situation can be considered either classically, i.e. disk scattering means performing specular reflections at the disk boundaries, or quantum mechanically, in which case scattering at the disks implies Dirichlet boundary conditions at the disk boundaries. 
This seemingly simple physical system is the paradigmatic model for an open chaotic system and has been introduced by Ikawa \cite{Ika88} and Gaspard-Rice \cite{GR89,GR89a, GR89b}. 
Let us summarize its key features:

\begin{itemize}
 \item Nearly all classical trajectories escape towards infinity making the system \emph{open}. Those trajectories that stay bounded both in positive and negative time form a compact \emph{fractal} set. The dynamics on this fractal set is \emph{uniformly hyperbolic}. 
 \item Openness of the quantum scattering problem is reflected by an exponential decay of the wave function \cite{Ika88, NZ09}. Furthermore the system possesses a mathematically well defined theory of \emph{quantum resonances}: 
 The resolvent of the Laplacian with Dirichlet boundary conditions admits a meromorphic continuation to $\mathbb{C}$ \cite{DZ19}. Furthermore efficient numerical algorithms for the calculation of these quantum resonances \cite{GR89b, Wir99} as well as the associated resonant states \cite{WBK+14} have been developed. 
 They rely on explicit expressions for the quantum scattering matrix developed in terms of incoming and outgoing radial waves.
 \item The classical dynamics has recently been shown to fit into the framework of open hyperbolic systems with compact trapped set that allows for a rigorous definition of Ruelle resonances, invariant Ruelle distributions and weighted zeta functions \cite{KSW21}. 
 Again efficient numerical methods (ranging under the name cycle expansion in the physics literature \cite{CE89}, see also \cite{JP02, Bor14, BW16, BPSW21} for related mathematical works) for the calculation of dynamical zeta functions for 3-disk systems as well as their poles and zeros are available.
 \item We have recently demonstrated \cite[Appendix~A]{BSW21} that the cycle expansion also allows for an efficient numerical calculation of the residues of \emph{weighted} zeta functions and thus of the invariant Ruelle distributions\footnote{Note that the applicability of the cycle expansion for this task had already been predicted by Eckhardt et al., but we are not aware of any published rigorous demonstrations that this is indeed a fact}. 
 \item Finally -- but from the physics perspective most importantly -- 3-disk systems allow for \emph{experimental realizations}, for example via microwave scattering systems \cite{LRPS99, PWB+12, BWP+13}. 
\end{itemize}

Let us explain more in depth the objects that we calculate: On the classical side the dynamics combines free propagation with uniform speed with specular reflections at the disk boundaries.
Even though the latter are certainly not smooth, it has recently been shown \cite{KSW21} that one can construct a smooth model for this dynamics that satisfies all requirements for an application of the mathematical results on weighted zeta functions and their residues discussed above. 
As suggested by the comparison between weighted and semiclassical zeta functions (see Section~\ref{sec:semiclassical}) we lift the flow to the bundle $|E_s|^{-1/2}$. 
Furthermore we have to take into account the Maslov indices of closed trajectories (in contrast to the example of a geodesic flow presented previously where these indices vanished). 

For a quantum billiard with Dirichlet boundary conditions, the Maslov index of a closed geodesic is twice the number $n_\gamma$ of boundary reflections making the phase contribution of the Maslov index equal to $\mathrm{i}\pi n_\gamma$ (see e.g.~\cite{CE89}). 
In order to take this into account we consider a complex line bundle $L^{\mathrm{ref}}$ that is glued at the disk boundaries in such a way that every reflection contributes with a factor of $-1$ (cf.~\cite[Remark~5.12]{KSW21}). 
We can then multiply this line bundle with the $1/2$-density bundle and lift the flow to the bundle $|E_s|^{-1/2}\otimes L^{\mathrm{ref}}$. 
The corresponding weighted zeta function then becomes
\[
 Z_a^{\mathrm{Ruelle}}(\lambda) = \sum_{\gamma} \left( \frac{\exp\left(-\lambda L_\gamma -\mathrm{i}\pi n_\gamma\right) |\Lambda_\gamma|^{1/2} } {\vert \det(\mathrm{id} - \mathcal{P}_\gamma) \vert} \int_{\gamma^\#} a \right).
\]

\begin{figure}[H]
	\includegraphics[scale=0.8]{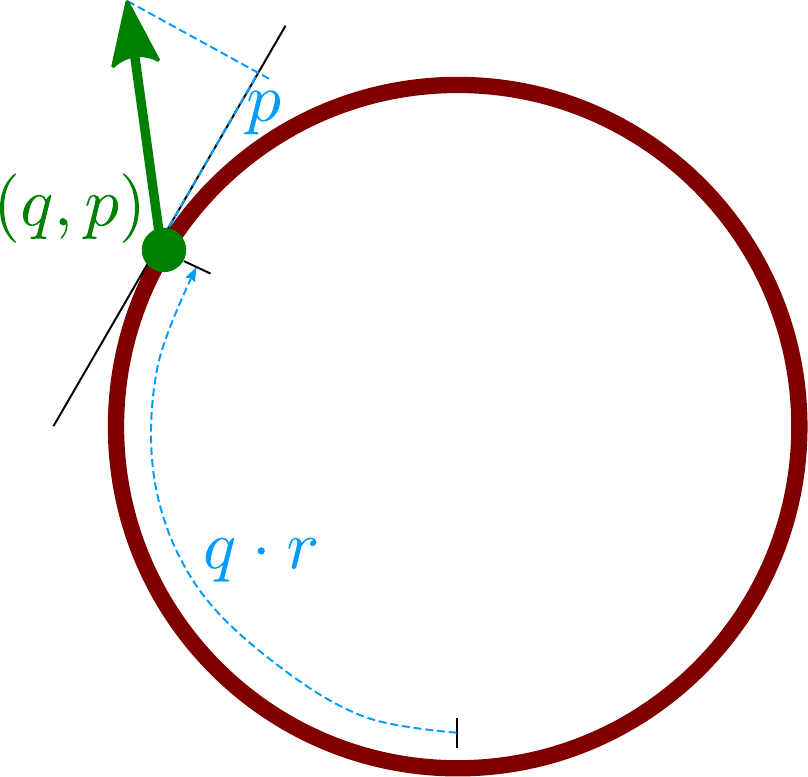}
	\caption{Illustration of the Poincar\'{e} section of boundary reflections. This 2-dimensional hypersurface in the energy shell $\{(x, y, \vec{v})\in \R^2\times \R^2 \,|\, \vert \vec{v}\vert = 1 \}$ is conveniently parameterized by Birkhoff coordinates $(q, p)$, where $q$ denotes the arclength distance along the disk boundary from some fixed base point (in units of the disk radius $r$) and $p$ is the projection of the velocity onto the tangent line to the disk.}
	\label{fig2}
\end{figure}

We can apply cycle expansion methods to calculate the poles and residues of this weighted zeta function. 
Actually we are not only interested in the value of $\mc T_{\lambda_0}(a)$ for a fixed weight $a$, but rather in the invariant Ruelle distribution $a\mapsto \mathcal{T}_{\lambda_0}(a)$ itself. 
In order to visualize this distribution we choose the following approach: 
First we restrict the distribution to the Poincar\'e section of reflections at the disk boundary.\footnote{This restriction is in fact possible as the wavefront set of $\mc T_{\lambda_0}$ is contained in $E_s^*\oplus E_u^*$. We can thus restrict it to any hypersurface that is transversal to the flow.}
On the disk boundaries we use Birkhoff coordinates $(q,p)\in [-\pi,\pi]\times[-1,1]$, see Figure \ref{fig2}.

We then consider the family of Gaussians $\varphi_{q,p;\sigma}$ centered at $q,p$ with variance $\sigma$ and calculate numerically the smooth function given by
\begin{equation*}
t_{\lambda_0, \sigma}: (q,p) \longmapsto \mathcal T_{\lambda_0}(\varphi_{q,p;\sigma}) .
\end{equation*}
Numerical plots of $t_{\lambda_0, \sigma}$ can be found in the bottom row of Figures~\ref{fig:dr6} and \ref{fig:dr3}. Note that $\langle t_{\lambda_0, \sigma}, a\rangle \rightarrow \mathcal{T}_{\lambda_0}(a)$ as $\sigma\rightarrow 0$, making $t_{\lambda_0, \sigma}$ a suitable smooth approximation for the generally quite singular (think of $\delta_0$) generalized function $\mathcal{T}_{\lambda_0}$.

On the quantum side we are interested in eigenfunctions $\Delta\vert n \rangle = k_n^2\vert n\rangle$ the distribution $a(q,p) \mapsto \langle n\vert \mathrm{Op}_{1/k_n}(a) \vert n\rangle$. Note that for $\mathrm{Op}_h$ being the quantization procedure commonly known as \emph{Weyl quantization} we get 
\[
 \langle n\vert \mathrm{Op}_{1/k_n}(a) \vert n\rangle  = \int a(q,p) W_{\vert n\rangle}(q,p) \mathrm{d}q \mathrm{d}p ,
\]
with $W_{\vert n\rangle}(q,p)$ being the \emph{Wigner phase space distribution} of the quantum eigenstate $\vert n\rangle$. As we are interested in the semiclassical limit anyway we can equivalently take any quantum phase space distribution that is semiclassically equivalent to the Wigner distribution. It turns out that a choice of phase space distribution particularly well suited for numerical calculations in the 3-disk system as well as for visualization purposes is given by the Poincar\'e-Husimi distribution. See \cite{WBK+14} for details and for a description of the concrete numerical algorithm. Again we restrict to the Poincar\'e section of boundary reflections and use Birkhoff coordinates on the disk boundaries. As we are dealing with an open quantum system, we have to distinguish between left and right resonant states $\langle n_L\vert$ and $\vert n_R\rangle$. We consider a combination of both resonant states as suggested in \cite{ECS09} which finally leads us to numerically calculate the smooth map
\[
 h_{k_n}: [-\pi,\pi]\times[-1,1]\ni(q,p) \longmapsto \langle c_{q,p; k_n}\vert n_R\rangle \langle n_L \vert c_{q,p; k_n}\rangle ,
\]
where $c_{q,p; k_n}$ is a \emph{coherent state} supported on the Poincar\'e section of boundary reflections. For additional details, in particular concerning the reduction to the Poincar\'e section, we refer to \cite[Section~2]{WBK+14}. Plots of the Poincar\'e-Husimi distributions for certain resonances of three disk systems can be found in the middle row of Figures~\ref{fig:dr6} and \ref{fig:dr3}.

Note that both, the classical and the quantum computations allow for symmetry reductions \cite{GR89b,CE93, BW16}. In the sequel we will present all results in the $A_2$ symmetry reduction that has the advantage of admitting a physical interpretation as the reduction to the fundamental domain of the billiard, see the green wedge in Figure~\ref{fig:3disk}.

\begin{figure}[H]
	\includegraphics[scale=0.75, clip={0mm 0cm 5cm 0cm}]{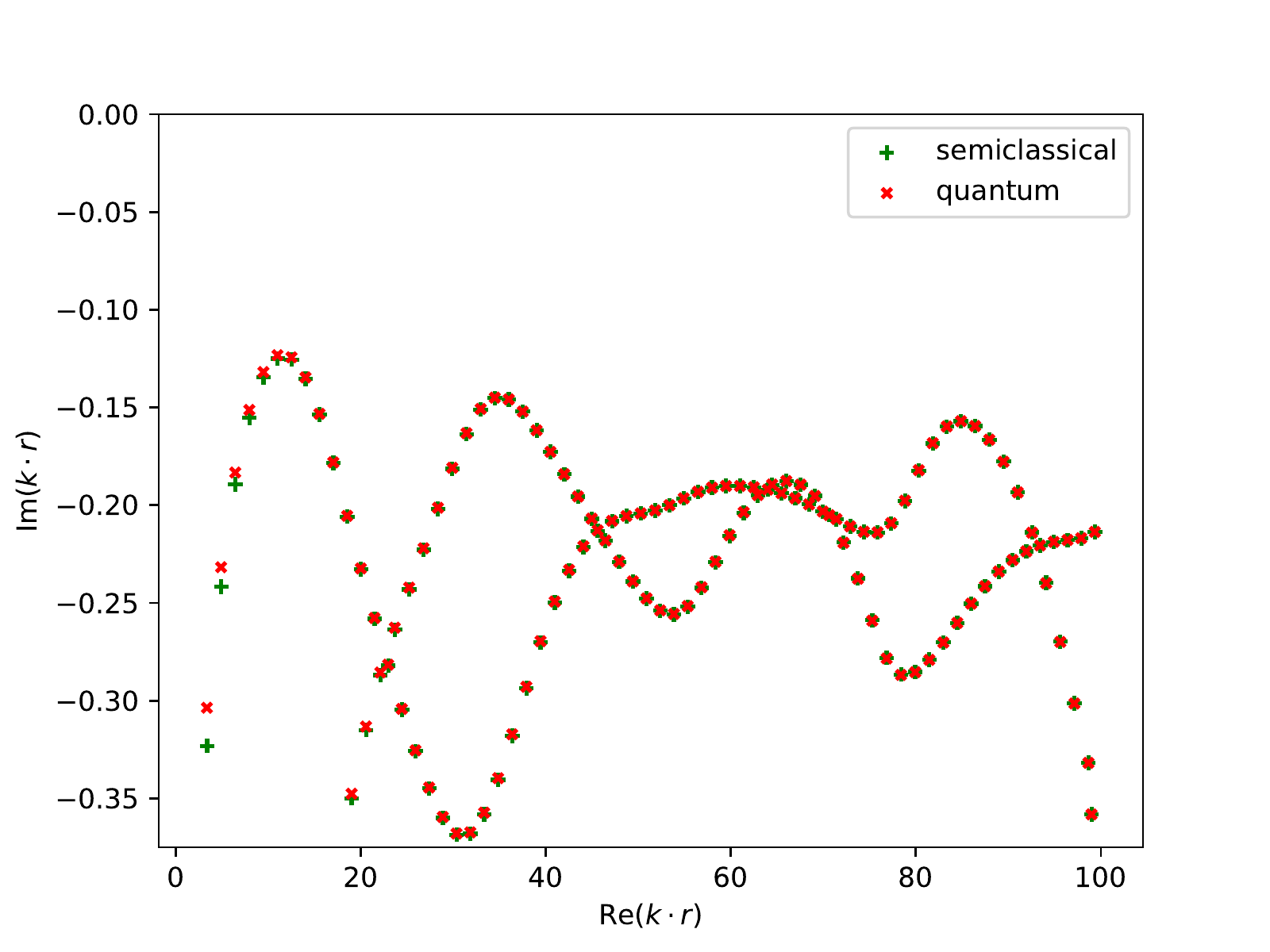}
	\caption{Comparison of numerically calculated Ruelle resonances of the billiard flow lifted to $|E_s|^{-1/2}\otimes L^\mathrm{ref}$ ({\color{mygreen} green $+$}) and quantum mechanical resonances ({\color{red} red $\times$}) for a 3-disk system with $d / r = 6$ (both computations done in symmetry reduction w.r.t. the $A_2$ representation of the symmetry group $C_{3v}$). We observe that for sufficiently large real part Ruelle and quantum resonances do indeed coincide with high precision.}
	\label{fig:compare_res}
\end{figure}

\begin{figure}[h]
	\includegraphics[scale=0.42]{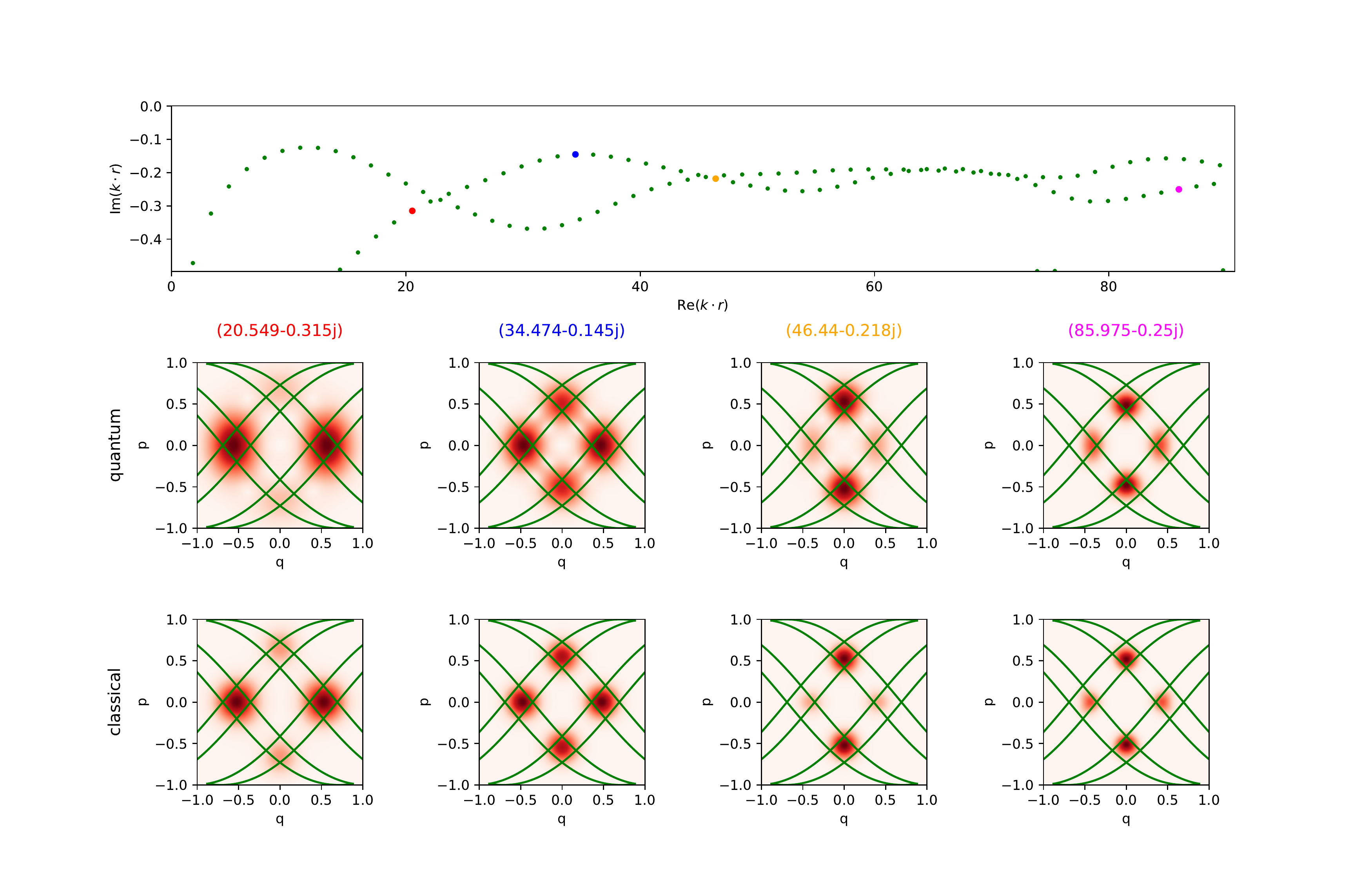}
	\caption{Numerical calculation of the resonances, Poincar\'e-Husimi distributions and invariant Ruelle distributions for a 3-disk system with $d / r = 6$. The first row shows a portion of the Ruelle resonances of the billiard flow lifted to $|E_s|^{-1/2}\otimes L^\mathrm{ref}$ (which coincide in high precision with the quantum resonances, see Figure~\ref{fig:compare_res}). The second and third rows compare the absolute values of the functions $h_{k_n}(q, p)$ (quantum) and $t_{k_n, 1/\mathrm{Re}(k_n)}(q, p)$ (classical). Every column corresponds to one specific resonance whose numerical value is indicated at the top of the column and is highlighted in the same color as the corresponding point in the resonance plot. All calculations were done in $A_2$ symmetry reduction.}
	\label{fig:dr6}
\end{figure}

\begin{figure}[h]
	\includegraphics[scale=0.42]{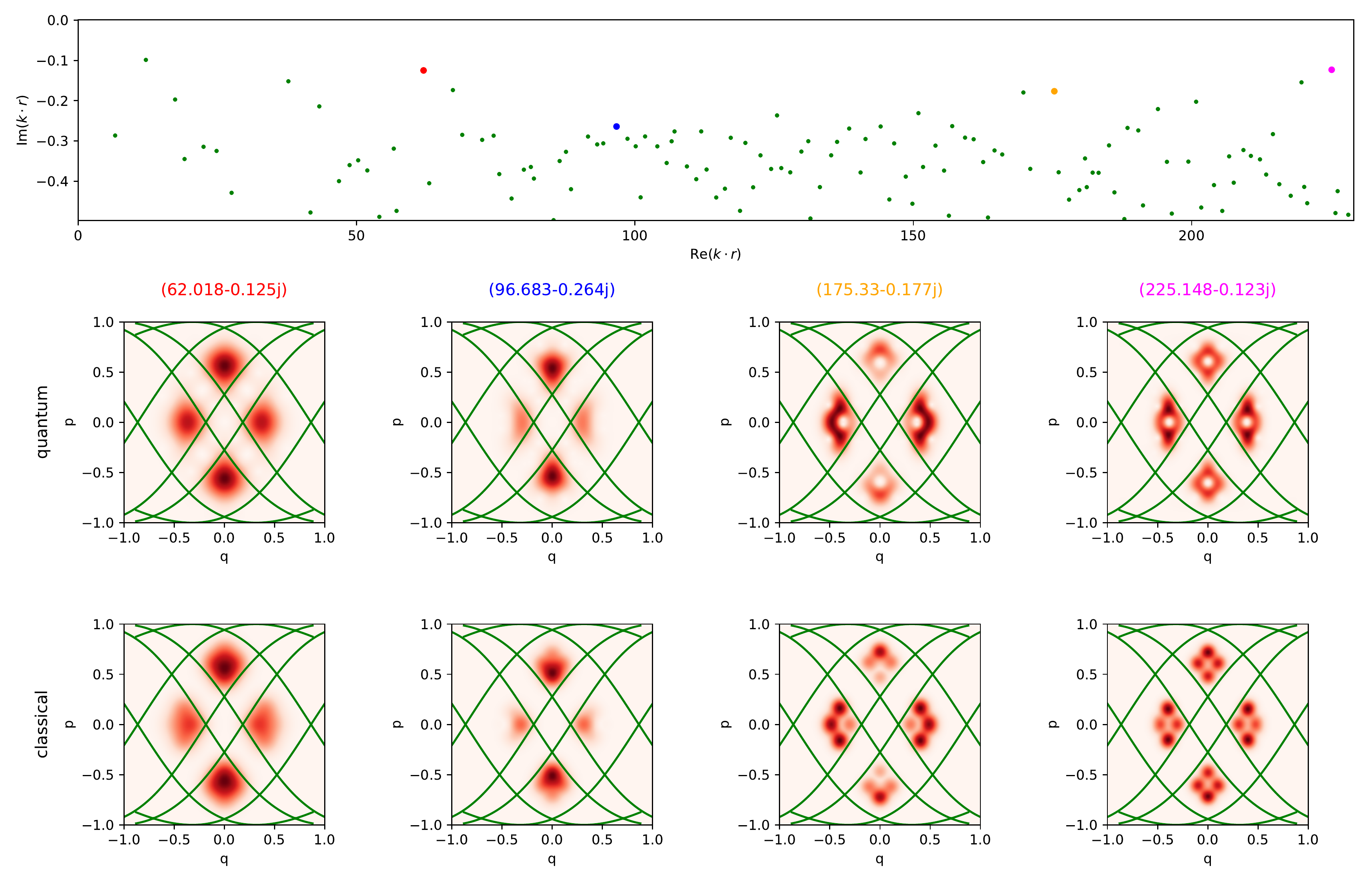}
	\caption{A second numerical calculation of semiclassical resonances, Poincar\'e-Husimi distributions and invariant Ruelle distributions, this time for a 3-disk system with $d / r = 3$. The first row shows a portion of the semiclassical resonance set. As in the previous figure the second and third rows compare the absolute values of the functions $h_{k_n}(q, p)$ (quantum) and $t_{k_n, 1/\mathrm{Re}(k_n)}(q, p)$ (classical), with every column corresponding to a specific color-coded resonance. Again all calculations were done in $A_2$ symmetry reduction.}
	\label{fig:dr3}
\end{figure}

Let us now turn to the comparison of the (classical) invariant Ruelle distributions and the (quantum) Poincar\'e-Husimi distributions:
As a Poincar\'e-Husimi distribution can be associated to a quantum resonance (i.e. pole of the quantum scattering matrix) and a invariant Ruelle distribution to a Ruelle resonance, one first of all needs a one-to-one correspondence between these two spectral quantities.
Figure~\ref{fig:compare_res} depicts the quantum resonances ({\color{red}red $\times$}) and the Ruelle resonances of the classical dynamics lifted to $|E_s|^{-1/2}\otimes L^\mathrm{ref}$ ({\color{mygreen} green $+$}). Note that for the Ruelle resonances we have turned the complex plane by 90 degrees and plotted $k_n=i\lambda$ such that the resonances match.  One sees that any quantum resonance exhibits a unique nearby Ruelle resonance and vice versa. 
Apart from the first few resonances also their value coincides with very good precision. In view of the development \eqref{eq:compare_zetas} this observation is a mere reformulation of the good agreement between quantum and semiclassical resonances which dates back to Cvitanovic-Eckhardt \cite{CE89} (see also \cite{Wir99}).

This one-to-one correspondence now allows us to compare, for a given resonance, the Poincar\'e-Husimi distribution and the invariant Ruelle distribution.
We present numerical data for two different 3-disk systems, first with the most commonly studied parameter $d/r=6$ (see Figure~\ref{fig:dr6}) and second for a more closed system with $d/r=3$ (see Figure~\ref{fig:dr3}). 
In both cases we have chosen four different resonances (highlighted in the resonance plot that is shown in the first row) and compare for each resonance the (quantum) Poincar\'e-Husimi distribution (second row) with the invariant Ruelle distribution (third row). Note that the invariant Ruelle distribution additionally depends on the smoothing parameter/Gaussian width $\sigma$. 
In order to take into account the increasing quantum resolution for higher frequencies we made the variable choice $\sigma = 1/\textup{Re}(k_n)$. Comparing the two plots we observe a qualitatively good agreement between the two distributions. While the exact form of the distributions differ, the localization on different regions of the fractal trapped set coincides very well for all plotted resonances. 
This degree of agreement between the two kinds of distributions has been observed over the whole plotted resonance spectrum. 

\begin{figure}[H]
	\includegraphics[scale=0.75]{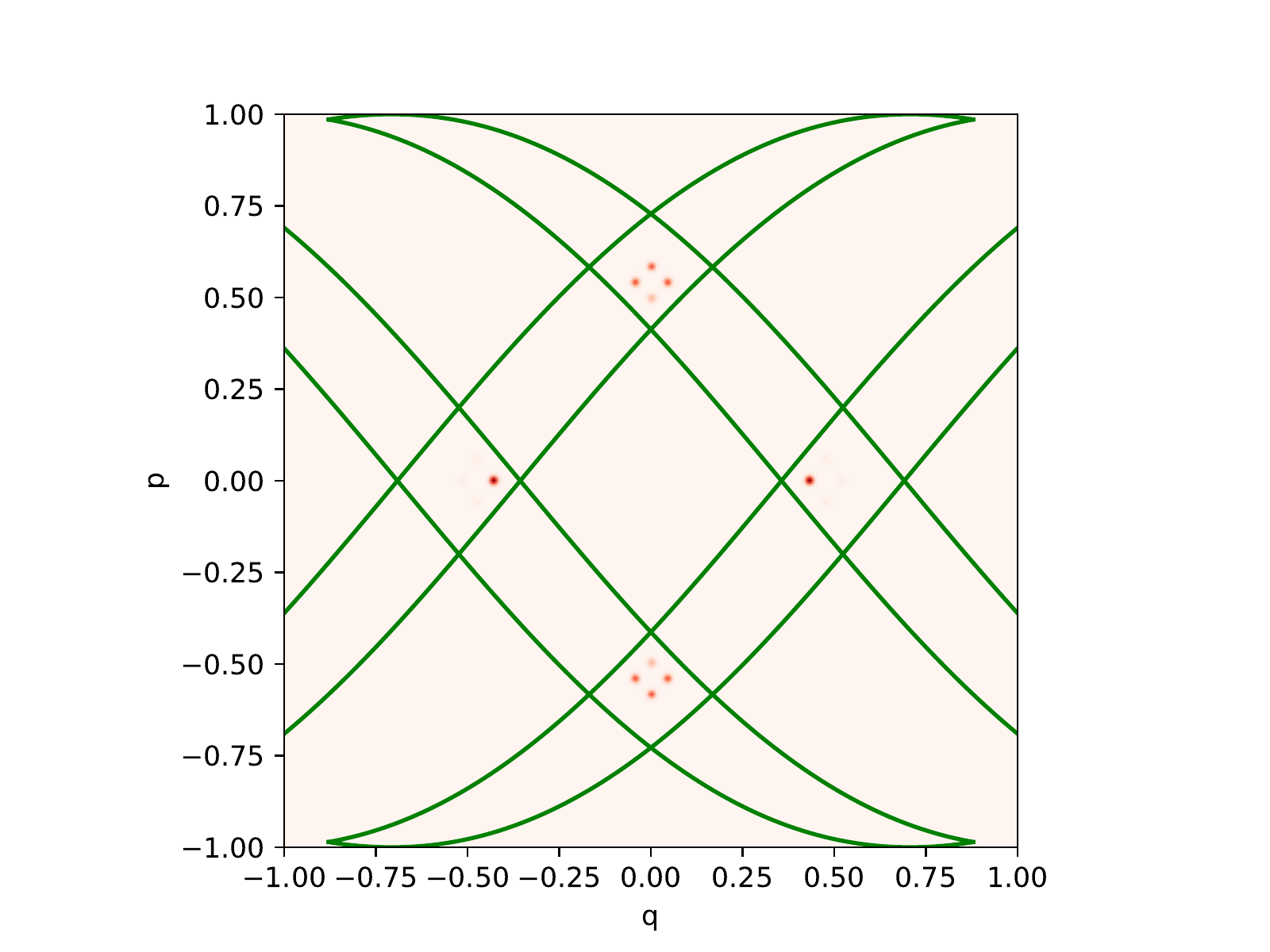}
	\caption{Numerical calculation of the absolute value of the function $t_{k_n, 1/\mathrm{Re}(k_n)}(q, p)$ for the specific resonance $k_n = 10000.983 - 0.207\mathrm{i}$ for a 3-disk system with $d/r=6$.}
	\label{fig:high_freq}
\end{figure}

Let us end the discussion of the numerical experiments by pointing out that the calculation of the invariant Ruelle densities is numerically much more feasible then the full quantum mechanical computation of Poincar\'e-Husimi distributions.
On a standard desktop PC and for a 3-disk system with $d/r=6$ the algorithm presented in \cite{WBK+14} for the full quantum computation is restricted to a frequency region of $\textup{Re}(kr)\lesssim 300$ due to the increasing size of the scattering matrix. In contrast, the cycle expansion of the zeta functions easily converges for $\textup{Re}(kr)\sim 10^{4}$ and allows for the computation of the corresponding invariant Ruelle densities in this frequency domain (see Figure~\ref{fig:high_freq}). 

\section{Conlusion and Outlook}
In this article we have reviewed the notion of invariant Ruelle distributions that can be associated to any Ruelle resonance (Section~\ref{sec:Ruelle}). 
For a large class of uniformly hyperbolic dynamical systems these invariant Ruelle distributions can be rigorously expressed as residues of weighted zeta functions (see \eqref{eq:ruelle_residue}). 
The flexibility of lifting the hyperbolic dynamics to vector bundles allows us to generate weighted zeta functions that can be related to previously derived semiclassical zeta functions (Section~\ref{sec:semiclassical}). 
Already in the 90s semiclassical arguments from \cite{EFMW92} predicted that the residues of these semiclassical zeta function are approximately given by quantum matrix coefficients. 
By relating the semiclassical weighted zeta functions to the weighted zeta functions for Ruelle distribution the results of \cite{EFMW92} predict an approximate equality between quantum matrix coefficients and invariant Ruelle distributions (see \eqref{eq:compare_inv_ruelle_quant_matrix}). 
Based on a mathematically sound quantum-classical correspondence, we explained that the semiclassical predictions are nowadays rigorous mathematical theorems in the setting of closed surfaces of constant negative curvature or more generally closed locally symmetric spaces of rank one (see Section~\ref{sec:exact_qcc}). 
In Section~\ref{sec:numerics} we then turned to more general and physically more relevant systems, namely 3-disk scattering systems, that do not fit into the framework of locally symmetric spaces.
Here, the question of relating invariant Ruelle distributions to quantum phase space distributions is mathematically still open. 
However, we show that both quantities can be computed numerically and in our numerical results we find a good qualitative agreement between the invariant Ruelle distributions and quantum phase space distributions, which supports the semiclassical predictions. 
This might not be surprising from the point of view of semiclassical theoretical arguments. However from the point of view of the recent works of Faure and Tsujii \cite{FT21} this is not at all obvious: For manifolds of variable negative curvature they manage to relate the Ruelle resonances of the geodesic flow lifted to $|E_s|^{-1/2}$ to some quantum spectrum.
But this quantum operator is not the exact Laplacian but contains lower order corrections. 
A priori it is not at all clear that these lower order corrections have a negligible impact on the spectrum in the semiclassical limit. 

But are the presented numerical agreements not strong evidence for the validity of the semiclassical predictions \eqref{eq:sc_residue} and \eqref{eq:compare_inv_ruelle_quant_matrix} that make mathematically rigorous analysis superfluous? 
There are certain caveats when drawing such a conclusion: The 3-disk system does not fit into the framework of locally symmetric spaces, where a rigorous version of \eqref{eq:compare_inv_ruelle_quant_matrix} is expected from a mathematical point of view.\footnote{Note
that \Cref{thm:qcc} does in this form only hold for closed manifolds. We conjecture, however, that an analogous statement can be shown also for open systems in constant curvature such as convex-cocompact hyperbolic surfaces.}
But important aspects of the symmetric 3-disk systems studied in this article are dynamically very similar to a dynamics on constant curvature spaces: 
The ratio $\log(\Lambda_\gamma)/L_\gamma$ varies for all calculated geodesics up to order $12$ ($\sim 8000$ different geodesics) by $\pm 0.5\%$ and $\pm 3.5\%$ for the 3-disk systems with $d/r=6$ and $d/r=3$, respectively. 
Thus both systems have very strongly pinched Lyapunov exponents, whereas geodesic flows in constant negative curvature exhibit exact pinching, cf. \eqref{eq:constant_curvature}. 
The numerically observed good agreement could thus also be an artifact of the strong pinching for symmetric 3-disk systems instead of evidence for the validity of \eqref{eq:sc_residue} and \eqref{eq:compare_inv_ruelle_quant_matrix} for arbitrary hyperbolic dynamics.   
We therefore consider it worthwhile to perform further numerical experiments in particular for systems that do not have such a strong pinching of the Lyapunov exponents as the symmetric 3-disk systems do.

\textbf{Acknowledgements:}
We thank Fr\'ed\'eric Faure for numerous enlightening discussions. This work has received funding from the Deutsche Forschungsgemeinschaft (DFG) (Grant No. WE 6173/1-1 Emmy Noether group “Microlocal Methods for Hyperbolic Dynamics”) and by an individual grant of P.S. from the Studienstiftung des Deutschen Volkes.


\bibliographystyle{amsalpha}
\bibliography{JRbib}
\bigskip

\end{document}